\documentclass[aps,prl,superscriptaddress,twocolumn,showpacs]{revtex4-1}
\usepackage[utf8]{inputenc}
\usepackage[american,british]{babel}
\usepackage[T1]{fontenc}
\usepackage[pdftex]{graphicx}
\usepackage{xcolor}
\usepackage{dcolumn}
\usepackage{bm}
\usepackage{amsmath,amsthm,amssymb,mathrsfs,mathtools,amsfonts,braket}
\usepackage{verbatim}
\usepackage{epsfig}
\usepackage{color}
\usepackage{xfrac}
\usepackage{geometry}
\usepackage{hyperref}
\usepackage{bbold}
\usepackage{tensor}
\usepackage{ulem}

\usepackage{booktabs}

\DeclareMathOperator{\polylog}{polylog}

\renewcommand{\sim}{\thicksim}

\usepackage{slashed}

\usepackage{tikz}

\newcommand{\urmove}[0]{%
\begin{tikzpicture}[baseline=0.ex]%
\draw[thick, red] (0,0ex) -- (1.25ex,1.25ex);%
\draw[thick, red] (1.25ex,1.25ex) -- (2.5ex,0);%
\end{tikzpicture}%
}

\newcommand{\drmove}[0]{%
\begin{tikzpicture}[baseline=0.ex]%
\draw[thick, red] (0,1.25ex) -- (1.25ex,0);%
\draw[thick, red] (1.25ex,0) -- (2.5ex,1.25ex);%
\end{tikzpicture}%
}

\usepackage{hyperref}
\hypersetup{
    bookmarks=false,         
    unicode=false,          
    pdftoolbar=false,        
    pdfmenubar=true,        
    pdffitwindow=false,     
    pdfstartview={FitH},    
    pdftitle={},    
    pdfauthor={Authors},     
    pdfsubject={},   
    pdfcreator={},   
    pdfproducer={}, 
    pdfnewwindow=true,      
    colorlinks=true,       
    linkcolor=black,          
    citecolor=blue,        
    filecolor=magenta,      
    urlcolor=blue           
}

\graphicspath{{../imag/}}

\begin{document}

\preprint{APS/123-QED}

\title{Simulating Schwinger model dynamics with quasi-one-dimensional qubit arrays}

\author{Alessio Lerose}
\affiliation{The Rudolf Peierls Centre for Theoretical Physics, Oxford University, Oxford OX1 3NP, United Kingdom}


\begin{abstract} 
Real-time dynamics of the Schwinger model provide an effective description of quark confinement out of equilibrium, routinely employed to model hadronization processes in particle-physics event generators. 
\textit{Ab-initio} simulations of such non-perturbative processes are far beyond the reach of existing computational tools, and remain an outstanding open quest for quantum simulators to date. 
In this work we develop a general strategy to run Schwinger model dynamics on synthetic quantum spin lattices, such as neutral-atom or superconducting-qubit arrays. 
Our construction encodes the constrained fermionic and bosonic degrees of freedom of the model into the geometric shape of a magnetic interface. We show that global magnetic field patterns can drive coherent \mbox{quantum dynamics} of the interface equivalent to the lattice Schwinger Hamiltonian.
We rigorously establish that the optimal array required for simulating real-time wave packet collisions and string fragmentation processes with accuracy $\epsilon$ in the continuum field-theory limit, is a quasi-one-dimensional ribbon with polynomial length and polylogarithmic width in $\epsilon^{-1}$. 
We finally discuss a concrete advantageous implementation using a state-of-the-art dual-species Rydberg atom array.
This work opens up a path for near-term quantum simulators to address questions of immediate relevance to particle physics.
\end{abstract}

\maketitle

\noindent Quantum simulation with controllable many-body quantum systems promises to revolutionize our understanding of many properties of matter in and out of equilibrium~\cite{Bloch2012,Blatt2012,cirac2012goals,georgescu2014quantum,Preskill2018quantumcomputingin}.
Its scope potentially includes nuclear and particle physics~\cite{byrnes2006simulating,wiese2013ultracold,ZoharReview,dalmonte2016lattice,banuls2020simulating,bauer2023quantumsimulation,bauer2023quantum,PRXQuantum.5.037001}:  Simulating real-time dynamics of strongly interacting fields \textit{ab-initio} would offer invaluable insights on non-perturbative phenomena at low energy scales, such as the string-fragmentation and hadronization processes governing the late stages of relativistic hadron and nuclei collisions~\cite{skands2013introduction}. 
Such phenomena, far beyond the reach of available computational methods, are routinely treated phenomenologically~\cite{ANDERSSON198331,Lund} and remain poorly understood at a fundamental level.

Field theories of fundamental interactions such as Quantum Chromo-Dynamics (QCD) are described by Lagrangians in $(3+1)$-dimensional spacetime, possessing fermionic and bosonic fields constrained by gauge symmetry as well as a \mbox{wealth of internal} degrees of freedom (spin, color, flavor)~\cite{skands2013introduction}. While these features make direct quantum simulation approaches prohibitive for near-term technologies, simpler dimensionally reduced versions of these theories do provide valuable information~\cite{banuls2020simulating,ZoharReview,wiese2013ultracold,dalmonte2016lattice}. For instance, string-fragmentation processes are thought to be effectively described by $(1+1)$-dimensional theories~\cite{casher1979chromoelectric,ANDERSSON198331,Lund} akin to the Schwinger model~\cite{SchwingerModel}. 
This minimal gauge theory has long served as {testbed}, from non-perturbative phenomena such as quark confinement and string breaking~\cite{HAMER1982413,buyens2016confinement,PichlerMontangeroQLMTensorNetwork}, chiral symmetry breaking~\cite{Banks1976,HAMER1982413,sachs2010finite}, quantum phase transitions induced by a topological angle~\cite{Coleman1976239,HAMER1982413}, to non-perturbative techniques such as tensor networks~\cite{byrnes2002density,banuls2013mass,banuls2016chiral,buyens2016hamiltonian,buyens2016confinement,PichlerMontangeroQLMTensorNetwork,banuls2017density,zapp2017tensor,Belyansky2024high,papaefstathiou2024real}.
Implementing controllable Schwinger model dynamics arguably represents the first milestone along the path leading quantum simulators to discoveries in high-energy physics.

In spite of its simplicity compared to QCD, experimental realization of the Schwinger model still faces the outstanding challenge of local gauge symmetry, which constrains interactions between fermionic matter and bosonic fields.
Various proposals addressed this challenge within digital~\cite{byrnes2006simulating,jordan2012quantum,jordan2014quantum,TAGLIACOZZO2013160,tagliacozzo2013simulation,mezzacapo2015non-abelian,zohar2017digital,klco2018quantum,lu2019simulations,arrighi2020quantum,klco2020su2,Shaw2020quantumalgorithms,jong2022quantum,nguyen2022digital,angelides2023first} or analog~\cite{zohar2011confinement,banerjee2012atomic,zohar2012simulating,Hauke2013b,zohar2013cold,zohar2013quantum,Banerjee2013,
wiese2013ultracold,marcos2013superconducting,stannigel2014,zou2014progress,Notarnicola_2015,yang2016analog,kuno2017quantum,gonzalez2017quantum,zache2018quantum,RICO2018466,notarnicola2020real,davoudi2020towards,luo2020framework,halimeh2022tuning,andrade2022engineering,aidelsburger2022cold} approaches, leading to the realization of pioneering digital simulations with trapped ions~\cite{ExpPaper,Muschik_2017} and superconducting-qubit processors~\cite{farrell2024quantum}, and of 
analog simulations of a simplified lattice gauge theory --- a $U(1)$ Quantum Link Model (QLM)~\cite{Chandrasekharan1997} with the gauge field represented by a spin-$1/2$ degree of freedom --- with one-dimensional Rydberg-atom arrays~\cite{bernien2017probing,SuraceRydberg} and optical lattices~\cite{Yang2020observation,zhao2022thermalization,zhang2306observation}. 
By construction, however, spin-$1/2$ QLM only provides a qualitative description of the continuum field-theory limit in a restricted range of parameters~\cite{Coleman1976239,CHANDRASEKHARAN1999739,schlittgen2001low,BROWER2004149,banerjee2012atomic, SuraceRydberg,zache2022toward}.
Proposals of scalable analog simulations of the \textit{bona fide} Schwinger model remain relatively scarce~\cite{kapit2011optical,kuhn2014quantum,yang2016analog,KASPER2016742,kasper2017implementing,Zache_2018,mil2020scalable,davoudi2021toward, Belyansky2024high,batini2024particle} and have not been experimentally realized yet.

In this work we construct a new strategy, readily applicable to synthetic short-range-interacting quantum spin lattices, such as neutral atom~\cite{Saffmann2010,kaufman2021quantum,barredo2016atom,endres2016atom,labuhn2016tunable,ebadi2021quantum,semeghini2021probing,bluvstein2022quantum,sheng2022defect,singh2023dual} or superconducting qubit~\cite{Kjaergaard2020Superconducting,arute2019quantum,kim2023evidence,wu2021strong,mi2022time} arrays. 

\medskip

\textit{Model.}
We consider  Quantum Electro-Dynamics in two space-time dimensions, known as Schwinger model~\cite{SchwingerModel}, described by the Lagrangian ($\hbar = c = 1$)
\begin{equation}
\label{eq_lagrangian}
\mathcal{L}=\bar{\psi}\Big(i\gamma^\mu({\partial}_\mu -q {A}_\mu)-m \Big)\psi -\frac{1}{4}F_{\mu\nu}F^{\mu\nu},
\end{equation}
 where $\mu,\nu=0,1$ label spacetime directions, $A_\mu$ and $F_{\mu\nu}=\partial_\mu A_\nu-\partial_\mu A_\nu$ encode the gauge potential and field, respectively, $\psi$ is a Dirac field with mass $m$ and charge $q$, and $\gamma^{0,1}$ are Dirac matrices. 
 We work with the Hamiltonian quantization of  Eq.~\eqref{eq_lagrangian} in the temporal gauge $A_0\equiv0$, whereby the potential $A^1$ and field $E=\dot A^1$ are canonically conjugated. 
 The Kogut-Susskind lattice Hamiltonian reads~\cite{KogutSusskindFormulation,Banks1976}
\begin{equation}
\label{eq_latticeschwinger}
\begin{split}
   \hat{H}_{\rm{Schwinger}} = 
   & -\frac i {2a} \sum_{j=1}^{2L-1} \left(\hat{c}^\dag_{j-1} 
   \hat U_{j-\frac12} 
   \hat{c}^{\mathstrut}_{j} - \textnormal{H.c.}\right) \\
   & -m \sum_{j=0}^{2L-1} (-1)^{j}\hat{c}_j^\dag\hat{c}^{\mathstrut}_j \\
   & + a \frac{q^2}2 \sum_{j=1}^{2L-1} \left(\hat L_{j-\frac12}- \frac{\vartheta}{2\pi}\right)^{\mathstrut 2}  ,
   \end{split}
\end{equation}
where $j=0,1,\dots,2L-1$ labels the sites of a discretized one-dimensional lattice with spacing $a$; the two components of $\psi$ are described by a single canonical Fermi field $\hat{c}_j$, i.e. $\hat \psi(x=2l a) = ({\hat c_{2l}}/{\sqrt{a}}, {\hat c_{2l+1}}/{\sqrt{a}})$; the parallel transporter $\hat U_{j+\frac12}= e^{-i a q  \hat A^1(x=ja)} $ {acts as} ladder operator for the dimensionless {electric field} $ \hat L_{j+\frac12}= \hat E\big(x=ja\big) / q $, i.e. $[\hat{L}_{i+\frac12}, \hat U_{j+\frac12}] =   \delta_{i,j} \hat U_{j+\frac12}$, realizing a chain of quantum rotors on the lattice bonds.
The additional parameter~$\vartheta$, known as \textit{topological angle}, physically represents a static electric field~\cite{Coleman1976239}.
The continuum limit is obtained as $a\to0$ and $L\to\infty$, with fixed physical volume $2La=\ell$.

\begin{figure}
    \centering
    \includegraphics[width=0.47\textwidth]{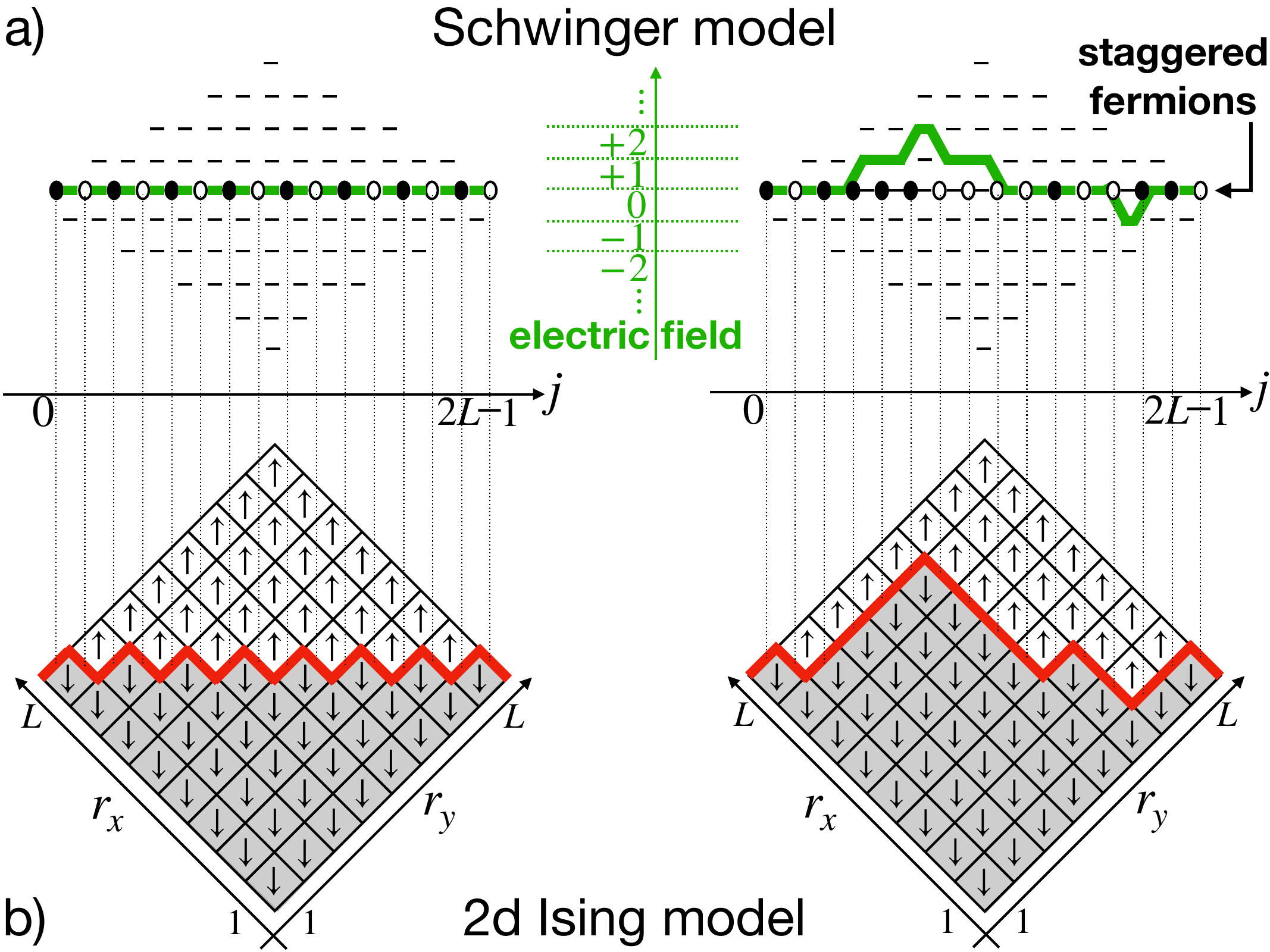}
    \caption{
    {\it Hilbert-space encoding in magnetic interfaces.}
        \textit{a)} Two examples of lattice Schwinger-model configurations, identified by a staggered-fermion occupation string $\{n_j=0,1\}_{j=0}^{2L-1}$ (white and black dots) and an electric field $\{l_{j-\frac12}=\dots,-2,-1,0,1,2,\dots\}_{j=1}^{2L-1}$ (green height profile), constrained by Gauss law $l_{j+\frac12}-l_{j-\frac12}=n_j-\frac{1+(-1)^{j}}{2}$ and charge neutrality $l_{-\frac12}= l_{2L+\frac12}=0$. The state on the left is the bare vacuum (strong-coupling ground state~\cite{Banks1976}).
        \textit{b)} Minimal-length Ising interfaces  encode Schwinger-model configurations, via the one-to-one correspondence highlighted by the vertical projection. 
    }
    \label{fig1}
\end{figure}

Assuming open boundary conditions, we define a computational basis labelled by fermion occupation numbers  and electric field values. A basis state $|\Psi\rangle=|n_0,l_{\frac12},n_1,l_{\frac32},n_2,\dots,n_{2L-2},l_{2L-\frac32},n_{2L-1}\rangle$ obeys $\hat c^\dagger_j \hat c_j |\Psi\rangle=n_j|\Psi\rangle$ and $\hat L_{j+\frac12} |\Psi\rangle=l_{j+\frac12}|\Psi\rangle$, with $n_j=0,1$ and $l_{j+\frac12}=\dots,-3,-2,-1,0,1,2,3,\dots$, as illustrated in Fig.~\ref{fig1}a.
The dimensionless charge content of site $j$ is $\hat Q_j = \hat{c}_j^\dagger \hat{c}^{\mathstrut}_j-\frac{1+(-1)^{j}}{2}$, i.e., a fermion [hole] at an odd [even] site represents a particle [antiparticle] with charge $+q$ [$-q$]; global charge neutrality  is given by the half-filling condition $\sum_{j=0}^{2L-1} \hat c^\dagger_j \hat c_j=L$. 
The generators of local $U(1)$ gauge transformations,  
$\hat G_j= \hat L_{j+\frac12}-\hat L_{j-\frac12}- \hat Q_j $, define superselection sectors. 
We will focus on the sector spanned by charge-neutral basis states 
obeying the discretized Gauss law 
$\hat G_j\Ket{\Psi}=0$, i.e., $n_j=l_{j+\frac12}-l_{j-\frac12}+\frac{1+(-1)^{j}}{2}$ for all $j$, as well as $l_{-\frac12}=l_{2L-\frac12}=0$. 
These constraints make labels redundant: One can express the local electric field as total charge to its left,
or, oppositely, the local charge as discrete derivative of the electric field profile.
The two elimination procedures lead, respectively, to a matter-only representation with long-range Coulomb interactions~\cite{Banks1976}, or a gauge-field-only representation with local constraints~\cite{ZoharFermionsToHCB,ZoharRemovingFermions,SuraceRydberg,LeroseSuraceQuasilocalization,borla2020gauging,notarnicola2020real,zache2022toward}.

\begin{table*}
\centering
\begin{tabular}{cc}
\toprule
2d Ising model \hspace{0.25cm} & Interface-encoded Hamiltonian \\
\midrule
$\hat H_{\rm{Ising}}= - J \sum_{\langle \mathbf{r},\mathbf{r'}\rangle} \hat\sigma^z_\mathbf{r} \hat\sigma^z_\mathbf{r'}$ \hspace{0.25cm} & ---  
\\
\midrule
$- g \sum_\mathbf{r} \hat \sigma^y_\mathbf{r}$ 
\hspace{0.25cm}& $- i g \sum_{j} \Big(\hat{c}^\dag_{j-1} 
   \hat U_{j-\frac12} \hat{c}^{\mathstrut}_{j} - \textnormal{H.c.}\Big)$  \\
$- h \sum_\mathbf{r} \hat \sigma^z_\mathbf{r}$  
\hspace{0.25cm}& $+2h\sum_{j} \hat L_{j-\frac12}$ 
 \\
$-h^\prime \sum_{\mathbf{r}} \Big( -L -2 + (r_x+r_y)+\frac{1+(-1)^{r_x+r_y}}2  \Big) \, \hat\sigma^z_{\mathbf{r}}$ 
\hspace{0.25cm}& $+2h^\prime \sum_{j} \hat L_{j-\frac12}^{\mathstrut 2}$ 
 \\
$-\mu \sum_{\mathbf{r}} (-1)^{r_x+r_y} \, \hat\sigma^z_{\mathbf{r}}$
\hspace{0.25cm}& $-\mu \sum_{j} (-1)^{j}\hat{c}_j^\dag\hat{c}^{\mathstrut}_j$ \\
\bottomrule
\end{tabular}
\caption{
Dictionary of Hamiltonian interface-encoding for Ising magnetic  interfaces (see Fig.~\ref{fig1}).
}
\label{tab_summaryIsing}
\end{table*}

\medskip

\textit{Hilbert-space encoding in magnetic interfaces.}
The key idea of this work is to encode a basis state  $|\Psi\rangle$ as  configuration of a domain wall in a two-dimensional magnet.
To see this, consider an Ising model on a square lattice, $\hat H_{\rm{Ising}}= - J \sum_{\langle \mathbf{r},\mathbf{r'}\rangle} \hat\sigma^z_\mathbf{r} \hat\sigma^z_\mathbf{r'}$, where $\mathbf{r}=(r_x,r_y)$, $r_x,r_y=1,\dots,L$ label the lattice sites, $\hat\sigma^{x,y,z}_\mathbf{r}$ are Pauli matrices, and $J>0$ is a ferromagnetic coupling between nearest-neighbor~spins. 
The system has two degenerate bulk configurations that minimize energy, i.e., $\hat\sigma^z_\mathbf{r}=+1$ or $-1$ for all~$\mathbf{r}$.
If boundary conditions enforce domain walls, the ground-state manifold is spanned by the configurations that minimize the domain wall length. 
Choosing opposite boundary conditions for the top and bottom sides of the $45^\circ$-rotated lattice, as shown in Fig.~\ref{fig1}b, minimal-length domain walls correspond to left-to-right-directed paths. On the one hand, such paths are in one-to-one correspondence with unconstrained sequences composed of~$L$ north-east and~$L$ south-east moves, equivalent to configurations of a half-filled fermionic chain of $2L$ sites. On the other hand, such paths can be directly interpreted as constrained height profiles, equivalent to configurations of a constrained bosonic field on a dual chain of $2L-1$ sites. These correspondences, illustrated by Fig.~\ref{fig1}, realize an isomorphism between Hilbert spaces: The degrees of freedom of our target lower-dimensional theory~\eqref{eq_latticeschwinger} are encoded in the Ising model ground-state manifold. Specifically, the sequence of path segments corresponds to the fermion occupation string $\big\{n_j\big\}_{j=0}^{2L-1}$ via the identification \textit{north-east} $\leftrightarrow n_j=1$, \textit{south-east} $\leftrightarrow n_j=0$,  whereas the height profile corresponds to the electric field profile $\big\{l_{j-\frac12}\big\}_{j=1}^{2L-1}$ via \textit{height}$\leftrightarrow l_{j-\frac12}+\frac{(-)^j}2$~(Fig.~\ref{fig1}).

We aim at inducing the target quantum dynamics on the interface by devising perturbations $\delta \hat H_{1,2,\dots}$ to $\hat H_{\rm{Ising}}$ whose matrix elements within the ground-state block exactly reproduce those of Eq.~\eqref{eq_latticeschwinger}, i.e. $\sum_p \hat P_{\rm{GS}}\,\delta\hat H_p \hat P_{\rm{GS}} \leftrightarrow \hat H_{\rm{Schwinger}}$, where $\hat P_{\rm{GS}}$ projects on the the ground-state manifold. 
Let us first consider a transverse magnetic field, $\delta\hat H_1 = - g \sum_\mathbf{r} \hat \sigma^y_\mathbf{r}$. Within the ground-state manifold $\delta\hat H_1$ only acts non-trivially at domain-wall turning points, i.e., $\hat P_{\rm{GS}}\,\delta\hat H_1 \hat P_{\rm{GS}} = -ig\sum_j \big(\ket{\drmove}_{\; j+\frac12}\bra{\urmove}-\ket{\urmove}_{\; j+\frac12}\bra{\drmove}\big)$~\cite{balducci2022localization,balducci2023interface}. 
These matrix elements exchange two consecutive path segments, simultaneously lowering or raising the path's local height by one unit. This coincides with the action of the kinetic term of $\hat H_{\rm{Schwinger}}$, first line in Eq.~\eqref{eq_latticeschwinger}. Thus, $g\leftrightarrow \frac 1 {2a}$.
 The remaining diagonal terms in Eq.~\eqref{eq_latticeschwinger} can be engineered via suitable longitudinal magnetic fields. 
A uniform field $\delta\hat H_2  = - h \sum_\mathbf{r} \hat \sigma^z_\mathbf{r}$, for example, takes values proportional to the area delimited by the interface. 
In terms of the height profile, $\hat P_{\rm{GS}}\,\delta\hat H_2 \hat P_{\rm{GS}} \leftrightarrow 2h(L+\sum_{j=1}^{2L-1} \hat L_{j-\frac12})$. Neglecting the additive constant, comparison with Eq.~\eqref{eq_latticeschwinger} yields $h \leftrightarrow  -\frac {a q^2 \vartheta}{4\pi}$.
The electrostatic energy, proportional to $\sum_{j=1}^{2L-1} \hat L^2_{j+\frac12}$, can then be generated using a vertical \textit{gradient}: With reference to the choice of coordinates in Fig.~\ref{fig1}b, one can check that the field pattern $\delta\hat H_3=-h^\prime \sum_{\mathbf{r}} \big( -L -2 + (r_x+r_y)+\frac{1+(-1)^{r_x+r_y}}2  \big) \, \hat\sigma^z_{\mathbf{r}}$ does the job, with $h^\prime \leftrightarrow  a\frac {q^2}4$.
Lastly, the mass term in Eq.~\eqref{eq_latticeschwinger} can be generated via a staggered field, $\delta\hat H_4= -\mu \sum_{\mathbf{r}} (-1)^{r_x+r_y} \, \hat\sigma^z_{\mathbf{r}}$, with $\mu \leftrightarrow  m$.
The mapping is summarized in Table~\ref{tab_summaryIsing}.


\textit{Scaling of resources in the continuum limit.} 
The quantum simulation strategy described above gets around the notorious difficulty of gauge-symmetry constraints. This advantage seemingly comes with a quadratic overhead in the number $N=L^2$ of qubits, as the effective simulated Hilbert space is that of a half-filled fermionic chain of $2L$ sites. How much of this redundancy can we dispose of by allowing for a small wave-function infidelity~$\varepsilon$? In fact, almost all of it:
We will prove reduction of the array size down to $N\lesssim L\log(L/\varepsilon)$ [respectively $ L^{3/2} \log^{1/2}(L/\varepsilon)$] for dynamical simulations at finite energy [finite energy density] above the ground state. 
These results refine the general truncation error bounds of Ref.~\cite{Tong2022provablyaccurate}.

Let us first consider the ground state $|\Phi_{m,q,\vartheta}\rangle$  of~\eqref{eq_latticeschwinger}. 
We scrutinize the error incurred upon truncating interface fluctuations outside of a horizontal ribbon of width $2W$. Denoting by $\hat\Pi_{W}$ the projection onto the subspace of configurations localized in this ribbon, the infidelity $1-|\langle  \Phi_{m,q,\vartheta}|\hat\Pi_{ W}|\Phi_{m,q,\vartheta}\rangle|^2$ is quantified by the probability $\mathbb{P}_{m,q,\vartheta}(|\hat L_{j-\frac12}|> W \text{ for some }j)$ that a measurement of the electric field outputs somewhere an absolute value larger than $W$. 
Since the probability of electric-field large deviations increase monotonically upon reducing the coupling $q$, computation of this quantity for a \textit{free} Dirac field provides an upper bound.
Electric field statistics 
is equivalent to the charge full counting statistics, which can be studied analytically in free-fermion lattice models using established results on the entanglement spectrum~\cite{peschel2009reduced}. Using these ideas, we can prove the bound~\cite{SM} $\mathbb{P}_{m,q,\vartheta}(|\hat L_{j-\frac12}|> W \text{ for some }j) \le 2L \frac {\lambda^{W+1}}{1-\lambda}$, where $\lambda = \exp(-\frac{\pi^2}{2 \log (\xi/a)})$ and $\xi$ denotes the correlation length in physical units~\footnote{This derivation applies to any value of $\vartheta$ \textit{not} equal to an odd-integer multiple of~$\pi$ --- the only effect of $\vartheta$ being a finite shift of the average electric field $\langle\hat L\rangle \simeq \vartheta/2\pi \ll W$ in the continuum limit. For $\vartheta=\pi (\mod 2\pi)$, however, the system has a quantum critical point in the 2d Ising universality class~\cite{Coleman1976239}, located at $m/q\simeq0.3335$~\cite{byrnes2002density}. The corresponding divergence $\xi=\infty$ of the correlation length affects the result through the replacement $\xi \mapsto \ell$.}.
A logarithmic cutoff $W\simeq \log(L/\varepsilon)$ thus suffices to ensure a large wave-function fidelity~$|\langle  \Phi_{m,q,\vartheta}|\hat\Pi_{ W}|\Phi_{m,q,\vartheta}\rangle|^2 \ge e^{-\varepsilon}$. 
The same conclusion applies to states with finitely many particle wave packets~\cite{jordan2012quantum,Surace_2021,karpov2022spatiotemporal,su2024cold,bennewitz2024simulating,rigobello2021entanglement,milsted2022collisions,papaefstathiou2024real,Belyansky2024high,farrell2024quantum}.

\begin{figure}
    \centering
    \includegraphics[width=0.47\textwidth]{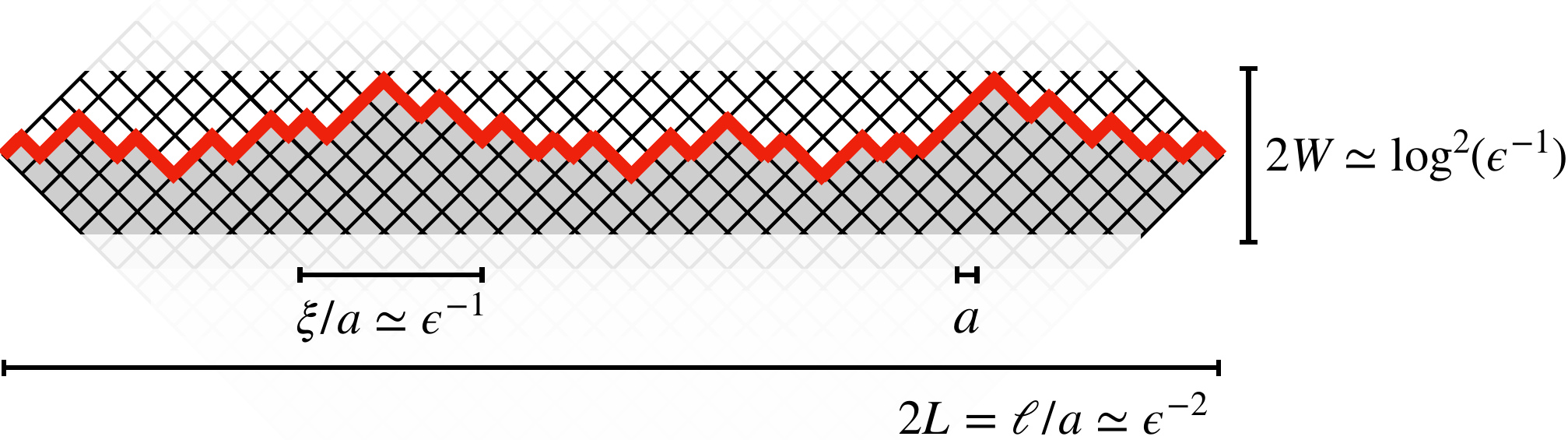}
    \caption{
    {\it Scaling of the qubit array size required for $\epsilon$-accurate simulations in the continuum field-theory limit.}  
    }
    \label{fig2}
\end{figure}

We now analyze the scaling of the required array size as the continuum limit is approached.
We consider the $\epsilon$-accurate determination of the distribution of particles emerging from the collision of meson wave packets~\cite{SchwingerScatteringReview,Belyansky2024high,papaefstathiou2024real} or from the string fragmentation of a receding electron-positron pair~\cite{ANDERSSON198331,Belyansky2024high,batini2024particle} as reference problems.
Errors in simulation results arise from putting the system in a finite volume (infrared cutoff~$\ell$), discretizing space (ultraviolet cutoff~$a\equiv\ell/2L$), and truncating electric-field fluctuations (local Hilbert space cutoff~$\pm W$).   
Firstly, a target resolution $\epsilon$ on particle momenta requires an infrared cutoff $\ell/\xi \simeq \epsilon^{-1}$.
Secondly, space discretization produces an error on local observables proportional to the lattice spacing (up to logarithmic corrections)~\cite{HamerSeriesExpansionSchwinger,byrnes2002density,banuls2013mass,kuhn2014quantum, banuls2016chiral, buyens2016hamiltonian,buyens2016confinement,banuls2017density}, leading to an ultraviolet cutoff $a/\xi \simeq \epsilon$. 
Lastly, as we target a small wave-function infidelity \textit{density} $\epsilon=\varepsilon/L$, the corresponding version of our electric-field truncation error bound 
$
\mathbb{P}_{m,q,\vartheta}(|\hat L_{j-\frac12}|> W ) \le  
      \log  (\xi/a) \, e^{- \frac W {\log (\xi/a) }} 
$~\cite{SM}
yields a cutoff $W \simeq \log^2(\epsilon^{-1})$.
The resulting scaling of the number of qubits is illustrated in Fig.~\ref{fig2}: %
\begin{equation}
\label{eq_resources}
    N_0  \,\lesssim\, \epsilon^{-2} \times \polylog \big(\epsilon^{-1}\big) \, .
\end{equation}

Equation~\eqref{eq_resources} applies to simulations at finite (volume-independent) energy above the ground state.
Concerning far-from-equilibrium dynamics, e.g. following a homogeneous quantum quench, electric field (or charge) fluctuations are expected to  significantly increase over time~\cite{Tong2022provablyaccurate}, before eventual saturation. 
It is natural to expect that the relevant maximum-entropy state, given by a Gibbs state with temperature $T>0$~\footnote{in our system, charge neutrality imposes a vanishing chemical potential}, provides an upper bound on fluctuations at all times. Similarly to $T=0$, finite-temperature charge fluctuations are maximized in the weak coupling limit. 
 Calculation for a free Dirac field~\cite{SM} shows that a wave-function infidelity smaller than~$\varepsilon$ is achieved through a cutoff $W\simeq \big(\sigma^2_T L\log(L/\varepsilon)\big)^{1/2}$, where $\sigma^2_T=\int_{-\pi/2}^{\pi/2}\frac{dk}{\pi} \frac 1 {1+\cosh(E_k/T)}$ and $E_k = \sqrt{m^2 + (\frac 1 a \sin^2k)^2} $.
Upon approaching the continuum  limit, 
we find the scaling of the array size~\footnote{Numerical constants, omitted for simplicity, carry the temperature dependence.}
\begin{equation}
\label{eq_resourcesT}
N_{T>0} \, \lesssim \,  \epsilon^{-5/2}\times \polylog \big( \epsilon^{-1}\big) \, .
\end{equation}

\begin{table*}
\begin{tabular}{cc}
\toprule
Neutral atom array & Interface-encoded Hamiltonian \\
\midrule
$
\hat H_{\rm{Ryd}} = \frac 1 2 \sum_{\mathbf{r},\mathbf{R}} \frac {V \delta_{\mathbf{R},\rm{even}}+V^\prime \delta_{\mathbf{R},\rm{odd}}} {|\mathbf{R}|^6} \hat n_{\mathbf{r}} \hat n_{\mathbf{r}+\mathbf{R}} -  \Delta  \sum_{\mathbf{r}}   \hat n_{\mathbf{r}}$
\hspace{0.3cm}
&
$2\sum_j \big(\frac {V^\prime}{125}-\frac V {64}  \big) \hat c_{j}^\dag \hat c_{j} \hat c_{j+1}^\dag \hat c_{j+1}+ \dots$ 
\\
\midrule
$ - \frac \Omega 2 \sum_{\mathbf{r}}    \hat\sigma^x_{\mathbf{r}} $
& $  -\frac14 \big(\frac{\Omega^2}{\tilde\Delta} + \frac{\Omega^2}{\tilde\Delta^\prime}\big) \sum_{j} \big(\hat{c}^\dag_{j-1}  \hat U_{j-\frac12} \hat{c}^{\mathstrut}_{j} + \textnormal{H.c.}\big) $  
\\
 $ -h \sum_{\mathbf{r}} \delta_{\mathbf{r},\rm{even}} \,\hat n_{\mathbf{r}} $ 
& $-h\sum_{j} \hat L_{j-\frac12}$ 
\\
$-h^\prime \sum_{\mathbf{r}} \delta_{\mathbf{r},\rm{even}} \Big(  r_y+\frac{1+(-1)^{r_y}}2  \Big) \, \hat n_{\mathbf{r}}$ 
& $ -h^\prime \sum_{j} \hat L_{j-\frac12}^{\mathstrut 2}$ 
\\
$ -\mu \sum_{\mathbf{r}}\delta_{\mathbf{r},\rm{even}} (-1)^{r_y} \,\hat n_{\mathbf{r}} $
& $+\frac \mu 2 \sum_{j} (-1)^{j}\hat{c}_j^\dag\hat{c}^{\mathstrut}_j$ 
\\
\bottomrule
\end{tabular}
\caption{
Dictionary of Hamiltonian interface-encoding for Rydberg-blockaded array (see Fig.~\ref{fig3}).
The mapping becomes exact for strictly nearest-neighbor interactions in the blockade regime. Interaction tails give rise to the weak spurious fermionic interactions reported in the right column, which can be strongly suppressed in a dual-species setup by tuning $V^\prime/V=125/64\simeq 1.953$. 
Consistency requires $0.5783 \, V \ll \Delta \ll 2.2953 \, V$~\cite{SM}. 
}
\label{tab_summaryRydberg}
\end{table*}

%
\begin{figure}
    \centering
    \includegraphics[width=0.47\textwidth]{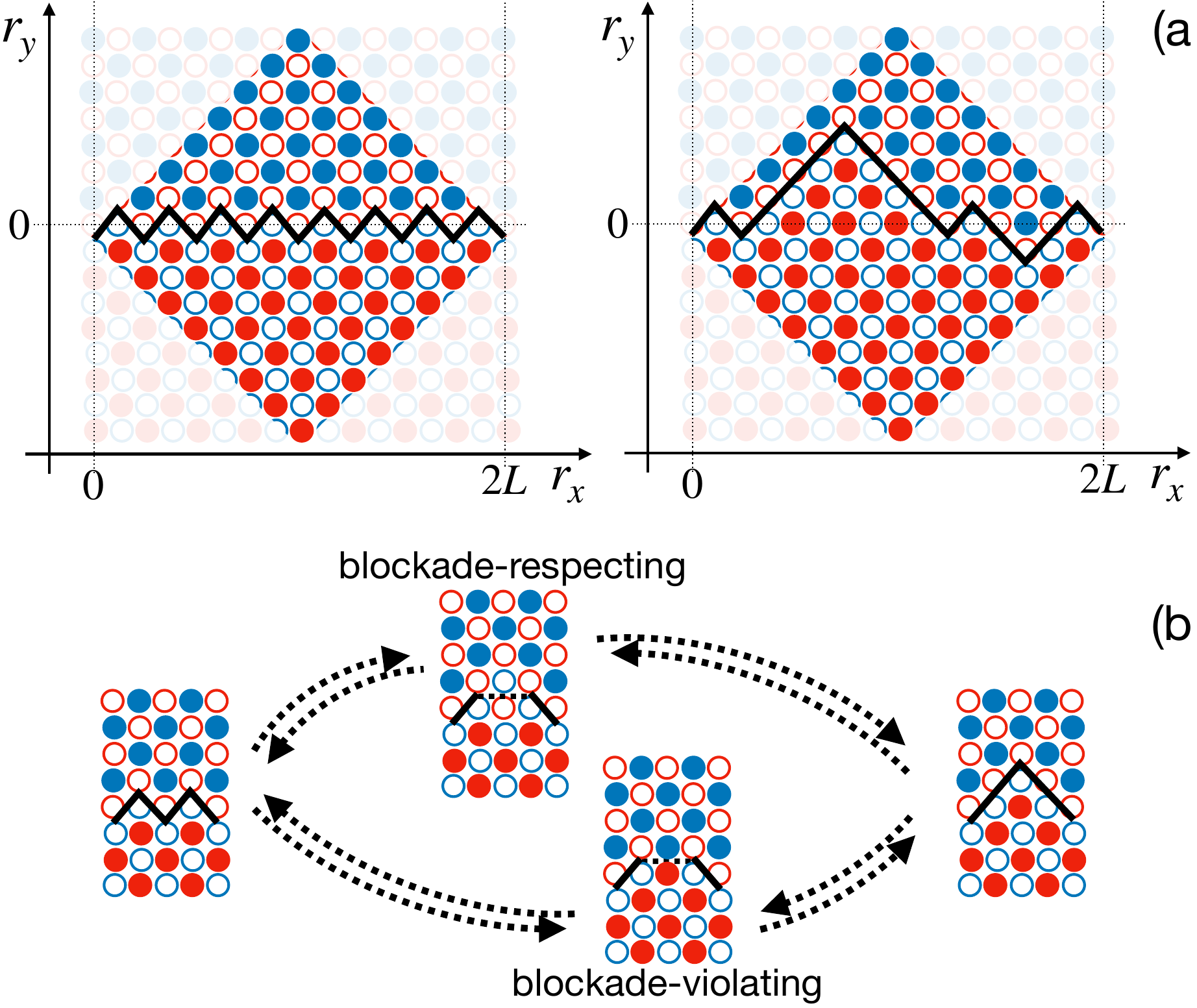}
    \caption{
    {\it Dual-species Rydberg-blockaded array implementation.}
        \textit{a)} 
        Checkerboard- and anti-checkerboard-ordered regions, composed of alternating ground (empty dots) and Rydberg (colored dots) atoms, are separated by Rydberg-blockade-respecting interfaces.
        Red and blue dots correspond to different atomic species, occupying the even and the odd sublattice, respectively.
        The two shown configurations realize the same Schwinger-model states as in Fig.~\ref{fig1}.
        \textit{b)} Second-order processes in~$\Omega$. 
    }
    \label{fig3}
\end{figure}

\textit{Implementation with  neutral atom arrays.}
We finally propose an example of concrete experimental implementation within the rapidly developing technology of neutral atom arrays~\cite{Saffmann2010,kaufman2021quantum,barredo2016atom,endres2016atom,labuhn2016tunable,semeghini2021probing,ebadi2021quantum,bluvstein2022quantum,sheng2022defect,singh2023dual}.
We consider atoms loaded in individual microtraps, configured in a 2d square lattice,  and coherently driven to a high-lying Rydberg state, where they experience strong van der Waals interactions.
The effective many-body Hamiltonian of the system can be written as
\begin{equation}
\label{eq_Rydberg}
    \hat H_{\rm{Ryd}} = \frac12 \sum_{ \mathbf{r}\neq\mathbf{r}^\prime} \frac {V}{|\mathbf{r}-\mathbf{r}^\prime|^6} \hat n_{\mathbf{r}} \hat n_{\mathbf{r}^\prime} - \frac \Omega 2 \sum_{\mathbf{r}}  \hat\sigma^x_{\mathbf{r}} -  \Delta \sum_{\mathbf{r}}   \hat n_{\mathbf{r}}  ,
\end{equation}
where the atom at site $\mathbf{r}=(r_x,r_y)$, $r_{x},r_y=1,\dots,L$ is described by a two-level Hilbert space, spanned by its ground $|g_\mathbf{r}\rangle$ and Rydberg $|r_\mathbf{r}\rangle$ states, acted upon by $\hat \sigma^{x}_\mathbf{r}\equiv |g_\mathbf{r}\rangle\langle r_\mathbf{r}|+ |r_\mathbf{r}\rangle\langle g_\mathbf{r}|$ and $\hat n_{\mathbf{r}}\equiv|r_\mathbf{r}\rangle\langle r_\mathbf{r}|$. Parameters $\Omega$ and $\Delta$ represent the effective Rabi frequency and detuning, respectively, of the laser-drive scheme, and the energy scale $V=\frac{C}{b^6}>0$ combines the relevant van der Waals coefficient~$C$ and the physical spacing~$b$ between the atoms.

While Eq.~\eqref{eq_Rydberg} can be directly  interpreted \mbox{as a 2d} quantum Ising model, the corresponding simulation protocol has drawbacks: 1) the sensitivity of interactions to the atoms' positional fluctuations  generates detrimental dephasing noise, and 2) van der Waals interaction tails introduce spurious terms in the encoded Hamiltonian which, while gauge-invariant, are not part of Eq.~\eqref{eq_latticeschwinger}.
Both issues can be largely overcome:\\ 
\indent 1) Dephasing noise is suppressed via the Rydberg-blockade effect~\cite{jaksch2000fast,lukin2001dipole,gaetan2009observation,urban2009observation}, which forbids simultaneous excitation of close-enough atoms. 
For $4 V_{nnn} \ll \Delta \ll 4 V_{nn}$, where $V_{nn}=V$ and $V_{nnn}=V/8$ are the two nearest neighbor couplings, an Ising-like symmetry-breaking phase compatible with nearest-neighbor Rydberg blockade occurs, characterized by alternating ground and Rydberg atoms in a checkerboard or anticheckerboard pattern~\cite{Samajdar2020complex,ebadi2021quantum}.
Provided $\Delta \ll V_{nn}+2V_{nnn}$, the blockade condition is also satisfied along checkerboard-anticheckerboard interfaces~\cite{SM}, which were recently experimentally realized~\cite{manovitz2024quantum}.
Minimal-length interfaces can be put in one-to-one correspondence with Schwinger-model configurations, as illustrated in Fig.~\ref{fig3}.\\ 
\indent 2) The weak spurious interface energy generated by third-nearest-neighbor and farther couplings, can be strongly suppressed by considering a dual-species array, inspired by the recent realization of configurable Rb-Cs arrays with individually controllable intraspecies and interspecies  interactions~\cite{singh2023mid,anand2024dual}.
Indeed, mismatch between third- and fourth-nearest-neighbor couplings gives rise to a configuration dependence of the interface energy, expressed by~\cite{SM} $ \hat H_{\rm{tail}}= (\frac V {64}- \frac V {125}) \sum_j  \big(\ket{\drmove}_{\; j}\bra{\drmove}+\ket{\urmove}_{\; j}\bra{\urmove}\big)$~\footnote{The Schwinger model subject to this perturbation is known as gauged Thirring or Schwinger-Thirring model. Physical properties are not critically affected by this perturbation~\cite{barros2019string}.}.
Its magnitude can be tuned by placing the two atomic species in the even and odd sublattice, respectively (see Fig.~\ref{fig3}), whereby $V$ in Eq.~\eqref{eq_Rydberg} gets replaced by $V \delta_{\mathbf{r}-\mathbf{r}^\prime \,\text{even}} + V^\prime \delta_{\mathbf{r}-\mathbf{r}^\prime \,\text{odd}}$. 
Tuning $V^\prime/V={125}/{64} \simeq 1.953 $ then leads to the desired suppression of~$\hat H_{\rm{tail}}$. 
Residual spurious terms generated by sixth-nearest-neighbor and farther couplings can be shown to be one order of magnitude weaker,
$ \hat H_{\rm{tail}} \simeq 0.0016 \, V $~\cite{SM}. 
Importantly, this suppression greatly enlarges the parameter range to approach the continuum limit~\footnote{With the notations of Table~\ref{tab_summaryRydberg},\\ $0.0016 \, V \ll h,h^\prime \ll \mu \ll \frac{\Omega^2}{\tilde\Delta} + \frac{\Omega^2}{\tilde\Delta^\prime} \ll V\approx \Delta$.}.

The encoding of Eq.~\eqref{eq_latticeschwinger} goes analogously to Table~\ref{tab_summaryIsing}.
The kinetic term is naturally generated by second-order processes in $\Omega$, which yield $\big(\frac{(\Omega/2)^2}{-\tilde\Delta+\tilde\epsilon} + \frac{(\Omega/2)^2}{-\tilde\Delta^\prime+\tilde\epsilon^\prime}\big) \sum_j \big(\ket{\drmove}_{\; j}\bra{\urmove}+\ket{\urmove}_{\; j}\bra{\drmove}\big)$~\footnote{This expression is equivalent to the one in Eq.~\eqref{eq_latticeschwinger} via the unitary transformation $\hat c_j \rightarrow e^{i(-1)^j \pi/4} \, \hat c_j$.}. Here, the denominators $\tilde{\Delta}=\Delta-0.3422 \, V$ and $\tilde{\Delta}^\prime=2.2953 \, V-\Delta$ are energy gaps associated with virtual excitations to blockade-respecting and blockade-violating configurations, respectively, illustrated in Fig.~\ref{fig3}b; the interface-configuration-dependent correction to the matrix element, accounted for by~$\tilde\epsilon$,~$\tilde\epsilon^\prime$, can be shown not to exceed $0.2\%$~\cite{SM}. 
The remaining diagonal terms can be generated via global laser patterns acting on a \textit{single} species, 
see Table~\ref{tab_summaryRydberg}~\footnote{We note that the denominators $\tilde\Delta$, $\tilde\Delta^\prime$ acquire slight shifts by $\mu$, $h$, $h^\prime$. This effect can be compensated, e.g., by implementing a corresponding weak spatial modulation of the Rabi frequency $\Omega \rightarrow \Omega_{\mathbf{r}}=\Omega \Big[1 + (-1)^{r_y}\mathcal{O}(\mu/V) + r_y\mathcal{O}(h^\prime/V) \Big] $.}. 

\medskip

\textit{\it Outlook.} 
We constructed a framework to simulate Schwinger model dynamics with the native degrees of freedom and interactions of qubit-array-based platforms. Our scheme circumvents the conventional difficulties associated with linearly growing Coulomb interactions or gauge-symmetry constraints.
Differently from dimensional reduction approaches~\cite{Chandrasekharan1997,BROWER2004149}, the extra spatial dimension is here designed to yield an \textit{exact} encoding of lattice fields (see also Refs.~\cite{zhang2018quantum,meurice2021theoretical}).

We expect that the reduction from the naive scaling $N=L^2\sim 1/\epsilon^4$ of the required number of qubits as a function of the target accuracy $\epsilon$, to the much more favorable ones in Eqs.~\eqref{eq_resources} and~\eqref{eq_resourcesT},  will prove critical for near-term experimental feasibility.
Furthermore, the bounds on electric-field fluctuations proven here provide a rigorous foundation for the use of either high-spin QLMs~\cite{Chandrasekharan1997} or other kinds of boson truncation~\cite{Notarnicola_2015,zache2023quantum} to systematically extrapolate observables in the continuum field-theory limit, both in matrix-product-state computations~\cite{buyens2016hamiltonian,buyens2016confinement,Belyansky2024high} and quantum simulation approaches~\cite{zache2022toward,Halimeh2022achievingquantum}.


\medskip

\begin{acknowledgments}
{\it Acknowledgments.}
I acknowledge funding through a Leverhulme-Peierls Fellowship at the University of Oxford.
I thank H. Bernien, M. Dalmonte, Z. Davoudi, H. Pichler, A. Prakash, P. Skands, F.  Surace, T. Zache, 
and E. Zohar for discussions and comments on the manuscript, as well as my co-authors of Refs.~\cite{SuraceRydberg},~\cite{balducci2022localization} for past collaborations which partially inspired this work.
\end{acknowledgments}


\bibliography{biblio}

\end{document}


\title{Supplemental Material: \\
Quantum simulation of the Schwinger model with quasi-one-dimensional qubit arrays}

\author{Alessio Lerose}
\affiliation{The Rudolf Peierls Centre for Theoretical Physics, Oxford University, Oxford OX1 3NP, United Kingdom}


\maketitle
\appendix

\section{
Bounds on electric-field fluctuations
}

As discussed in the main text, understanding the efficiency of quantum simulations of the Schwinger model requires to estimate a Hilbert-space truncation error in a thermal state, namely  
\begin{equation}
1-\Tr \left( \frac{e^{- \hat H_{\rm{Schwinger}}/T}}Z \hat\Pi_{ W} \right) \, ,
\end{equation}
where $\hat\Pi_{W}= \prod_j \hat\Pi_{j, W}$ simultaneously projects all local bosonic Hilbert spaces onto the subspaces where the electric field takes values smaller than $W$ in magnitude, i.e. $|\ell_{j-\frac12}| \le W$ for all $j$. 
This truncation error can be interpreted as the probability that measurements of the electric field in the thermal state take somewhere an absolute value larger than~$W$, i.e.,  $\mathbb{P}_{m,q,\vartheta}\Big(|\hat L_{j-\frac12}|> W \text{ for some }j\Big)$.  
By a simple union bound,
$\mathbb{P}_{m,q,\vartheta}\Big(|\hat L_{j-\frac12}|> W \text{ for some }j=1,\dots,2L-1\Big) \le \sum_{j=1}^{2L-1}\mathbb{P}_{m,q,\vartheta}\Big(|\hat L_{j-\frac12}|> W\Big)$. In the presence of open boundary conditions, fluctuations are translationally invariant in the bulk and suppressed near the left and right boundaries, hence we can bound them with their extent at the center of the system,
$\mathbb{P}_{m,q,\vartheta}\Big(|\hat L_{j-\frac12}|> W\Big) \le \mathbb{P}_{m,q,\vartheta}\Big(|\hat L_{L-\frac12}|> W\Big)$. Furthermore, a reduction of the electrostatic term $\propto q \sum_j \hat L_{j-\frac12}^2$ in the Schwinger model Hamiltonian can only monotonically increase electric-field fluctuations, which reach their maximum extent in the weak coupling limit $q\to0$. Putting everything together,
\begin{equation}
\label{eq_basicbound}
    1-\Tr \left( \frac{e^{- \hat H_{\rm{Schwinger}}/T}}Z \hat\Pi_{ W} \right) \le (2L-1) \, \mathbb{P}_{m,q\to0}\Big(|\hat L_{L-\frac12}|> W\Big) \, . 
\end{equation}
 In the considered limit of vanishing coupling $q\to0$, the Schwinger model reduces to a free Dirac field with mass $m$ (the value of $\vartheta$ becomes immaterial in this limit).
  Upon eliminating the electric field~\cite{Banks1976}, the exact fermionic lattice Hamiltonian for $q\to0$ reads
 \begin{equation}
\hat H_{\rm{Dirac}} = 
   -\frac i {2a} \sum_{j=1}^{2L-1} \left(\hat{c}^\dag_{j-1} 
\hat{c}^{\mathstrut}_{j} - \textnormal{H.c.}\right) 
    -m \sum_{j=0}^{2L-1} (-1)^{j}\hat{c}_j^\dag\hat{c}^{\mathstrut}_j .
 \end{equation}
Its bulk dispersion relation has two bands,
\begin{equation}
    E_k = \pm \sqrt{m^2 + \frac 1 {a^2} \sin^2 k}, \qquad -\frac \pi 2 \le k \le \frac \pi 2 \, ,
\end{equation}
which reduce to the standard result $E(p) = \pm \sqrt{m^2+p^2}$ in the continuum limit $k\equiv ap$, $a\to0$.

Recalling that $\hat L_{L-\frac12}=\sum_{j=0}^{L-1} \hat Q_j = \sum_{j=0}^{L-1} \hat c^\dagger_j \hat c_j-\frac L2$ for even $L$, our task of estimating the right-hand side of Eq.~\eqref{eq_basicbound} is equivalent to estimating the probability that, in a thermal state, the number of fermions $\hat{\mathcal{N}}_{\rm{left}}=\sum_{j=0}^{L-1} \hat c^\dagger_j \hat c_j$ contained the left half of the system has a large deviation from the  half-filling value $L/2$. 
Since $\hat{\mathcal{N}}=\hat{\mathcal{N}}_{\rm{left}}+\hat{\mathcal{N}}_{\rm{right}}=\sum_{j=0}^{2L-1} \hat c^\dagger_j \hat c_j$ commutes with $\hat H_{\rm{Dirac}}$, the reduced density matrix $\hat \rho_{\rm{left}} =  \Tr_{\rm{right}} \frac{e^{- \hat H_{\rm{Dirac}}/T}}Z$ commutes with~$\hat{\mathcal{N}}_{\rm{left}}$. Furthermore, as $\hat\rho_{\rm{left}}$ is Gaussian, we can express it as $\hat\rho_{\rm{left}} = \frac 1 \Xi \exp \left (- \sum_{\kappa=1}^L \beta_\kappa \hat d^\dagger_\kappa \hat d_\kappa \right) $, where $\{d_\kappa\}$ are rotated canonical fermions and  $\Xi=\prod_{\kappa=1}^L (1+e^{-\beta_\kappa})$,  and hence $\hat{\mathcal{N}}_{\rm{left}} = \sum_{\kappa=1}^L  \hat d^\dagger_\kappa \hat d_\kappa$. The statistics of the observable~$\hat{\mathcal{N}}_{\rm{left}}$ in the state~$\hat\rho_{\rm{left}}$  is thus equivalent to the sum of $L$ independent binary (Bernoulli) random variables with probabilities $p_\kappa = \frac1{1+e^{-\beta_\kappa}}$.
As a result of particle-hole (i.e., CP) symmetry, these probabilities can be arranged in such a way that $\frac12 \ge p_1\ge p_2 \ge\dots\ge p_{L/2}\ge0 $ and $p_{L/2+\kappa}=1-p_\kappa$ for all $\kappa=1,\dots,L/2$.
Thus, $\langle \hat{\mathcal{N}}_{\rm{left}} \rangle \equiv L/2$, and we inquire about large deviations from this average value.

\subsection{Charge full counting statistics in the ground state }

 We shall first consider the vacuum, $T\to0$. In this case, the numbers $\beta_\kappa$ determine the so-called entanglement spectrum.
 Analytical results on the entanglement spectrum of free-fermion lattice models~\cite{peschel2009reduced} show that in gapped systems the probabilities $p_\kappa$ decay exponentially in the thermodynamic limit,
 \begin{equation}
 p_\kappa \underset{\kappa\to\infty}{\thicksim} \lambda^\kappa \qquad \text{with} \qquad \lambda=e^{-\frac{\pi^2}2 \frac 1 {\log (\xi/a)} } \, ,
 \end{equation}
where $\xi/a$ is the ground-state correlation length in lattice-spacing units. 

Since the two subsequences $\kappa=1,\dots,L/2$ and $\kappa=L/2+1,\dots,L$ of Bernoulli variables approach a deterministic variable ($1$ and $0$, respectively) 
exponentially fast in $\kappa$, we can establish a direct bound on the probability of large deviations in this process. 

Let us first establish that the most likely subsequence of Bernoulli variables $\kappa=W+1,W+2,\dots,L/2$ (i.e. all outcomes $=1$) has a probability  $\mathbb{P}_{\rm{m. l. s.}}^{(W)}$ that remains finite as $L\to\infty$ and approaches $1$ as $W$ increases: 
\begin{multline}
    \log \mathbb{P}_{\rm{m. l. s.}}^{(W)} \; =  \; \sum_{\kappa=W+1}^{L/2} \log (1-\lambda^\kappa)
    = - \sum_{s=1}^\infty \frac 1 s \sum_{\kappa=W+1}^{L/2} \lambda^{s\kappa} \ge - \sum_{s=1}^\infty \frac 1 s \sum_{\kappa=W+1}^{\infty} \lambda^{s\kappa}
    = 
    \\
    = - \sum_{s=1}^\infty \frac 1 s \frac {\lambda^{s(W+1)}} {1-\lambda^{s}}
    \ge - \sum_{s=1}^\infty \frac 1 s \frac {\lambda^{s(W+1)}} {1-\lambda} = \frac{\log(1-\lambda^{W+1})}{1-\lambda} > -\infty \, ,
\end{multline}
whence
\begin{equation}
\label{eq_mostlikelysequence}
    \mathbb{P}_{\rm{m. l. s.}}^{(W)} \; \ge \;  \left(1-\lambda^{W+1}\right)^{\frac 1 {1-\lambda}} > 0 \, .
\end{equation}
Exactly the same statement applies to the subsequence $\kappa=L/2+W+1,L/2+W+2,\dots,L$.

Now, the occurrence of a large deviation event $|\mathcal{N}_{\rm{left}} - \frac L 2 | > W$ requires that at least one of the two subsequences, $\kappa=1,\dots,L/2$ or $\kappa=L/2+1,\dots,L$, has strictly more than $W$ unlikely Bernoulli outcomes. For definiteness, say the former has this property. Then, necessarily, in the subsequence $\kappa=W+1,W+2,\dots,L/2$ there is at least one unlikely outcome. For the probability of this event, the bound in Eq.~\eqref{eq_mostlikelysequence} applies: 
\begin{equation}
    \mathbb{P}\left(
\left|\mathcal{N}_{\rm{left}} - \frac L 2 \right| > W
    \right) \; \le \; 1-\mathbb{P}_{\rm{m. l. s.}}^{(W)} \; \le \; 1- \left(1-\lambda^{W+1}\right)^{\frac 1 {1-\lambda}} \; \underset{}{\thicksim} \; \frac{\lambda^{W+1}}{1-\lambda} \, .
\end{equation}

\label{app}

\subsection{Charge full counting statistics at finite temperature}

We shall now consider a thermal state with temperature $T>0$. In this case the half-chain reduced density matrix is also thermal, up to boundary corrections (negligible for large $L$). We can thus identify $\beta_\kappa \leftrightarrow E_k/T$.
As the variances of the $L$ independent Bernoulli variables are now all finite, the statistics of the sum can be handled with standard statistical-mechanics tools. 

Specifically, we compute the moment generating function of $\mathcal{N}_{\rm{left}}$,
\begin{equation}
    \langle e^{\alpha \hat{\mathcal{N}}_{\rm{left}}} \rangle = \prod_{k} \frac{1+e^{- E_k/T + \alpha}}{1+e^{- E_k/T }} \prod_{k} \frac{1+e^{+ E_k/T + \alpha}}{1+e^{+ E_k/T }} .
\end{equation}
and hence the scaled cumulant generating function,
\begin{equation}
\label{eq_scaledcumulant}
    \frac 1 {L} \log \langle e^{\alpha \hat{\mathcal{N}}_{\rm{left}}} \rangle  \; \underset{L\to\infty}{\thicksim} \; \frac \alpha 2 + 
    \int_{-\pi/2}^{\pi/2} \frac{dk}{2\pi} \log \bigg(\frac{\cosh(\alpha)+\cosh(E_k/T)}{1+\cosh(E_k/T)}\bigg) \; \equiv \; \Psi_T(\alpha) \, ,
\end{equation}
the Legendre transform of which, $\Phi_T(\mathcal{N})$, gives the desired large-deviation function
\begin{equation}
    \mathbb{P}\left(
\mathcal{N}_{\rm{left}} - \frac L 2 \approx W
    \right) \; \thicksim \; e^{-L \Phi(1/2 + W/L)} \, .
\end{equation}

For the purpose of our bound, it is sufficient to compute the size of typical fluctuations: Expansion of Eq.~\eqref{eq_scaledcumulant} to second order in $\alpha$,
\begin{equation}
    \Psi_T(\alpha) \thicksim  \alpha \frac12 + \frac {\alpha^2}2 \int_{-\pi/2}^{\pi/2} \frac{dk}{2\pi}  \, \frac{1}{1+\cosh(E_k/T)} + \mathcal{O}(\alpha^4)
\end{equation}
allows us to identify the scaled variance of the distribution
\begin{equation}
    \sigma^2_T = \lim_{L\to\infty} \frac 1 {L} \bigg\langle \left( \hat{\mathcal{N}}_{\rm{left}} - \frac L2 \right)^2 \bigg \rangle = \int_{-\pi/2}^{\pi/2} \frac{dk}{2\pi}  \, \frac{1}{1+\cosh(E_k/T)} \, .
\end{equation}
In the continuum limit $k=ap$, $a\to0$, this becomes
\begin{equation}
    \sigma^2_T  \; \underset{}{\thicksim} \; a \int_{-\infty}^{+\infty} \frac{dp}{2\pi}  \, \frac{1}{1+\cosh(\sqrt{m^2+p^2}/T)} \, .
\end{equation}

\section{
Effects of van-der-Waals interaction tails
}

The Rydberg-blockade-respecting interface encoding discussed in the main text becomes exact in the limit of strictly nearest-neighbor interactions.
Here we derive the effects of van-der-Waals interaction tails (beyond-nearest-neighbor interactions) on the interface potential and dynamics. We will work with the dual-species setup described in the main text. In all the calculations we will assume the form $V/r^6$ for the interaction strength and keep into account the first $8$ nearest neighbors only, neglecting nineth-nearest-neighbor interactions $V_{9n}=\frac V {\sqrt{18}^6}=\frac V {5832}$ and farther  ($r\ge \sqrt{18}$).

Let us first of all inquire the local bulk stability of checkerboard and anti-checkerboard configurations, as well as the local stability of Rydberg-blockade-respecting interfaces between them. 

De-exciting a Rydberg atom to its ground state in the bulk of a checkerboard state leads to an energy increase $\bar\Delta$ equal to the detuning $\Delta$, minus tail corrections,
\begin{equation}
\bar\Delta = 
    \Delta - 4 V \left[ 
    \frac 1 {\sqrt{2}^6} + \frac 1 {2^6} + \frac 1 {\sqrt{8}^6}
    + 2 \frac 1 {\sqrt{10}^6}
    \right] \simeq \Delta - 0.5783 \, V \, .
\end{equation}
On the other hand, exciting a ground-state atom in a checkerboard state leads, instead, to a large energy increase $\bar\Delta^\prime$ due to four-fold Rydberg-blockade-violations, plus tail corrections,
\begin{equation}
\bar\Delta^\prime = 
    -\Delta + 4 V^\prime \left[ 1+
    2 \frac 1 {\sqrt{5}^6} + \frac 1 {3^6}
    + 2 \frac 1 {\sqrt{13}^6}
    \right] \simeq  4.0734 \, V^\prime -\Delta  \, .
\end{equation}
Bulk stability of the checkerboard state requires positivity and largeness of both these energy gaps, i.e.,
\begin{equation}
\label{eq_bulkstability}
 0.5783 \, V \ll \Delta \ll 4.0734 \, V^\prime  \, , \qquad   \bar\Delta, \bar\Delta^\prime \gg \left|\frac \Omega 2\right| \, .
\end{equation}

In the main text we work in the regime where minimum-energy interfaces between checkerboard and anti-checkerboard regions are also Rydberg-blockade-respecting, i.e., composed of pairs of adjacent ground-state atoms (see Fig. 3 of the main text). In order for this to happen, exciting ground-state atoms along the interface has to lead to a large positive energy increase. This translates into the condition
\begin{equation}
\label{eq_interfacestability}
    \bar\Delta^{\prime\prime}
    = -\Delta + V^\prime + 2  \frac V {\sqrt{2}^6} +  \frac V {2^6} + 2  \frac {V^\prime} {\sqrt{5}^6} + \dots \gg \left|\frac \Omega 2\right| \, ,
\end{equation}
where the omitted terms ``\dots'' constitute a small configuration-dependent correction (see below). This interface-stability condition narrows down the bulk-stability upper bound~\eqref{eq_bulkstability} on $\Delta$.

 In summary, upon choosing $\Delta$ in the range determined by Eq.~\eqref{eq_bulkstability} (lower bound) and Eq.~\eqref{eq_interfacestability} (upper bound), checkerboard and anti-checkerboard patterns are locally stable bulk configurations, and blockade-respecting domain walls are locally stable interfaces.

\subsection{Spurious interactions}
 The various minimal-length domain-wall configurations are exactly degenerate in energy  only for strictly nearest-neighbor interactions. Beyond-nearest-neighbor couplings give rise to a weak configuration-dependent potential for the interface, which lifts the degeneracy and generates spurious interactions in the encoded one-dimensional Hamiltonian.

\begin{figure}
    \centering
    \includegraphics[width=0.6\textwidth]{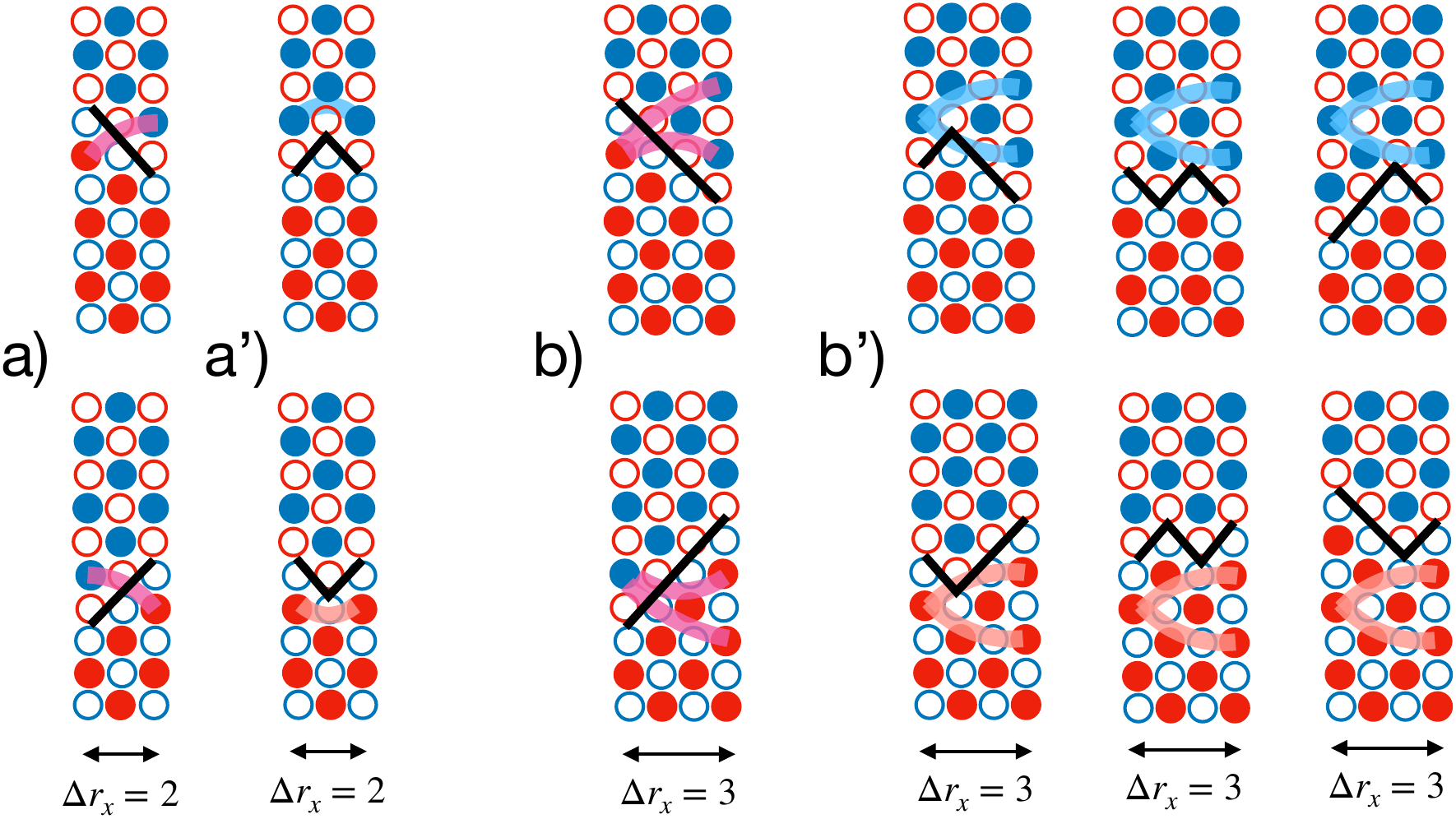}
    
    \caption{
    {\it Interface-configuration-dependent energy contributions from van der Waals interaction tails.}  
    }
    \label{fig_configdependence}
\end{figure}

Let us evaluate this configuration-dependent energy. Firstly, with reference to Fig. 3 of the main text, we observe that the interface has slope $\pm1$ everywhere. Interpreting the horizontal [vertical] lattice coordinate $r_x$ [$r_y$] as a fictitious ``time'' [``space''] coordinate, with ``speed of light'' equal to $1$, it is easy to see that couplings between atoms with ``causally-disconnected'' separation $|\Delta r_y| > |\Delta r_x|$ do not generate any configuration-dependent energy, whereas couplings between ``causally connected'' atoms with $|\Delta r_y|< |\Delta r_x|$ do.
The effect of such couplings can be organized in terms of increasing $\Delta r_x$.
The leading configuration-dependent energy contribution is given by pairs of atoms with $\Delta r_x=2$: Fig.~\ref{fig_configdependence}a,a' show that the energy difference between a straight segment of length 2 and a turning point is
\begin{equation}
    E(\uumove \text{ or }\ddmove) - E(\urmove\text{ or }\drmove) = - \frac{V}{2^6} + \frac {V^\prime}{\sqrt{5}^6}
\end{equation}
 Denoting by $\Circle$  [$\CIRCLE$] an empty  [occupied] fermionic site in the encoded 1d system, the encoded 1d Hamiltonian acquires 
 \begin{equation}
    \hat H_{\rm{tail}}= - \left(\frac{V}{2^6} - \frac {V^\prime} {\sqrt{5}^6}\right)
     \sum_j \left( \hat P _{\Circle_j\Circle_{j+1}} + \hat P_{\CIRCLE_j\CIRCLE_{j+1}} \right)
 \end{equation}
 where $\hat P$ denotes projection operators on the corresponding local configurations in subscripts.
 This term is equivalent to the weak fermionic density-density interaction reported in Table II of the main text.
In a dual-species setup where $V$ and $V^\prime$ can be independently tuned, this interaction can be fully suppressed by setting 
\begin{equation}
\label{eq_cancellation}
    V^\prime=\frac {125} {64} V \simeq 1.953 \, V \, .
\end{equation}

 With this choice of interspecies/intraspecies coupling ratio,  the leading configuration-dependent energy is given by pairs of atoms with $\Delta r_x=3$. As illustrated in Fig.~\ref{fig_configdependence}b,b', these couplings give rise to an energy mismatch between straight segments of length 3 and other configurations.
 The resulting spurious interaction in the encoded 1d Hamiltonian, 
\begin{equation}
\hat H_{\rm{tail}}=
\left(\frac{V^\prime}{3^6} + \frac {V^\prime} {\sqrt{13}^6}
      - 2 \frac {V} {\sqrt{10}^6}
      \right)
     \sum_j \left( \hat P _{\Circle_{j-1}\Circle_j\Circle_{j+1}} + \hat P_{\CIRCLE_{j-1}\CIRCLE_j\CIRCLE_{j+1}} \right)
     \simeq
     0.0016 \, V \sum_j \left( \hat P _{\Circle_{j-1}\Circle_j\Circle_{j+1}} + \hat P_{\CIRCLE_{j-1}\CIRCLE_j\CIRCLE_{j+1}} \right) ,
 \end{equation}
  is equivalent to a much weaker fermionic density-density interaction,  $ \hat H_{\rm{tail}}=0.0016 \, V \sum_j \big( \hat c_{j}^\dag \hat c_{j} \hat c_{j+1}^\dag \hat c_{j+1} + \frac12 \hat c_{j}^\dag \hat c_{j} \hat c_{j+2}^\dag \hat c_{j+2} \big)$. 

\subsection{Second-order interface dynamics}

Figure 3b of the main text illustrates the  second-order dynamical processes in the Rabi frequency $\Omega$. Here we sketch the computation of the energy gaps associated with the corresponding intermediate virtual states.

\begin{figure}
    \centering
    \includegraphics[height=0.22\textwidth]{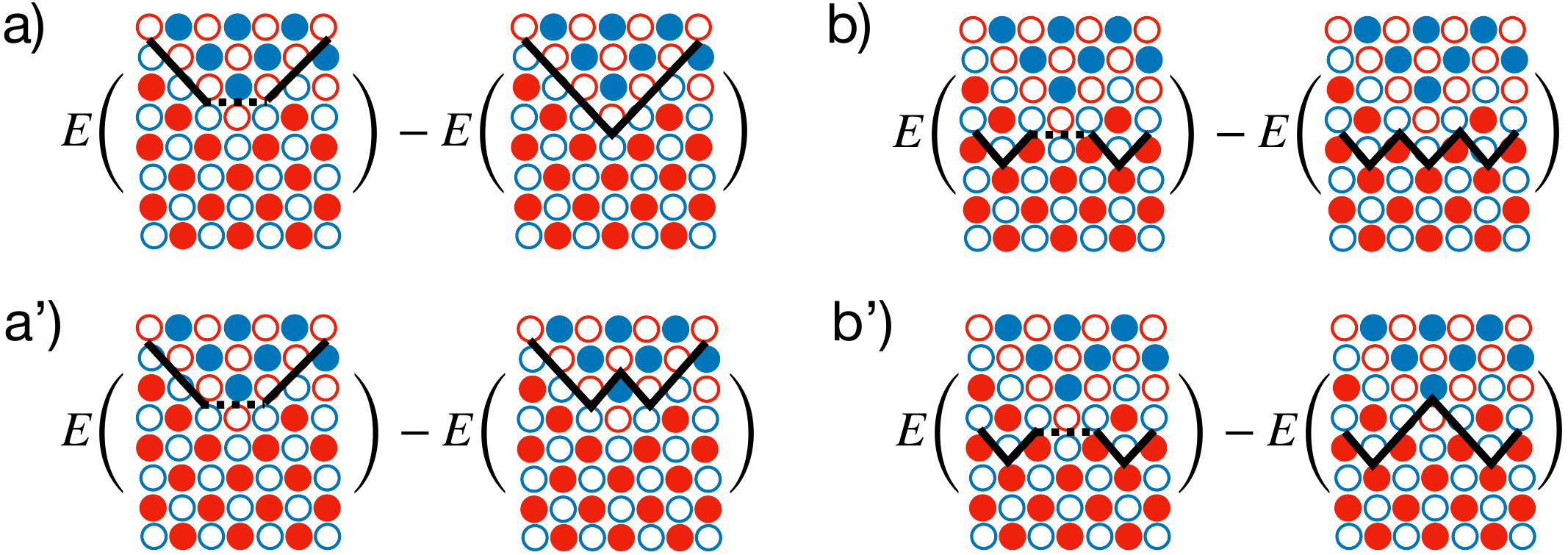}
\includegraphics[height=0.22\textwidth]{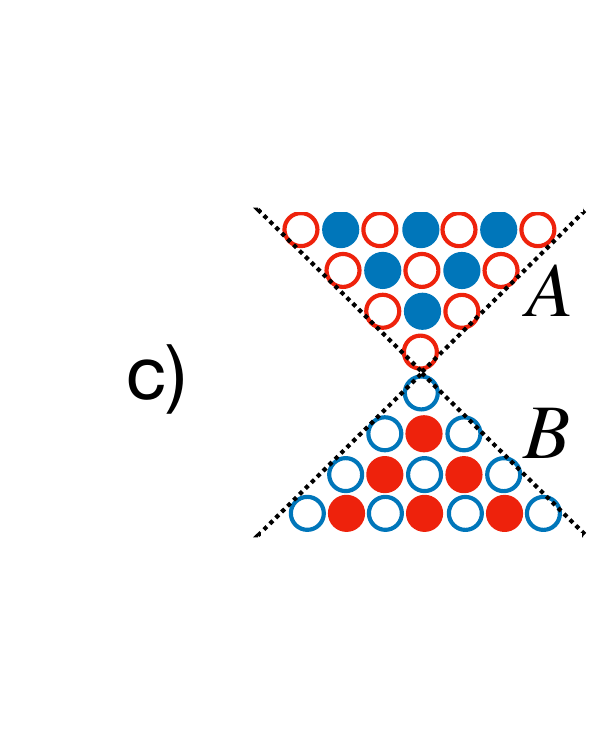}
    
    \caption{
    {\it Energy denominators for second-order blockade-respecting processes.}  
    }
    \label{fig_denominators}
\end{figure}

Let us consider the blockade-respecting process first. We evaluate the energy increase $\delta E$ associated with the virtual state for the specific interface configurations illustrated in Fig.~\ref{fig_denominators}a and b, for which the energy increase is smallest and largest, respectively. 
We obtain
\begin{equation}
    \delta E (\text{Fig. 2a}) =  \Delta + \delta E_A + \delta E_B  - 2 \frac{V^\prime}{\sqrt{5}^6} - 2 \frac{V^\prime}{{3}^6} - 2 \frac{V^\prime}{\sqrt{13}^6} \, ,
\end{equation}
\begin{equation}
    \delta E (\text{Fig. 2a'}) =  \Delta + \delta E_A + \delta E_B - 2 \frac{V}{2^6}  - 4 \frac{V}{\sqrt{10}^6} \, ,
\end{equation}
\begin{equation}
    \delta E (\text{Fig. 2b}) =  \Delta + \delta E_A + \delta E_B - 2 \frac{V}{2^6}  - 2 \frac{V}{\sqrt{10}^6}
    - 2 \frac{V^\prime}{\sqrt{13}^6} \, ,
\end{equation}
\begin{equation}
    \delta E (\text{Fig. 2b'}) =  \Delta + \delta E_A + \delta E_B - 2 \frac{V^\prime}{\sqrt{5}^6}  - 2 \frac{V}{\sqrt{10}^6}
    - 2 \frac{V^\prime}{\sqrt{13}^6} \, ,
\end{equation}
where we have defined the contributions to the energy increase from atoms in the ``causally disconnected'' regions A and B (see Fig.~\ref{fig_denominators}c),
\begin{equation}
\begin{split}
    \delta E_A =  & - 2 \frac{V}{\sqrt{2}^6} - \frac{V}{{2}^6} - 2 \frac{V}{2\sqrt{2}^6} - 2 \frac{V}{\sqrt{10}^6}   \\
    \delta E_B =  & 
    - 2 \frac{V^\prime}{\sqrt{5}^6} - \frac{V^\prime}{{3}^6} - \frac{V^\prime}{\sqrt{13}^6}
\end{split}
\end{equation}
 which are the same for all interface configurations (up to swapping A and B).
 Enforcing Eq.~\eqref{eq_cancellation}, the contribution of blockade-respecting second-order processes to the matrix element are
\begin{equation}
    \frac12 \bigg(\frac{\Omega}{2}\bigg)^2
    \bigg(
    \frac 1 {-\Delta + 0.3456 \, V}
    + \frac 1 {-\Delta + 0.3425 \, V}
    \bigg)
    \sum_j\big(\ket{\drmove}_{\; j}\bra{\urmove}+\ket{\urmove}_{\; j}\bra{\drmove}\big) \qquad \text{(for the configuration in Fig.~\ref{fig_denominators}a,a')}
\end{equation}
or
\begin{equation}
 \bigg(\frac{\Omega}{2}\bigg)^2
    \frac 1 {-\Delta + 0.3422 \, V}
\sum_j\big(\ket{\drmove}_{\; j}\bra{\urmove}+\ket{\urmove}_{\; j}\bra{\drmove}\big) \qquad \text{(for the configuration in Fig.~\ref{fig_denominators}b,b')} \, .
\end{equation}
We can thus rewrite it as
\begin{equation}
 \bigg(\frac{\Omega}{2}\bigg)^2
    \frac 1 {-\tilde\Delta + \tilde\epsilon}
\sum_j\big(\ket{\drmove}_{\; j}\bra{\urmove}+\ket{\urmove}_{\; j}\bra{\drmove}\big) \, ,
\end{equation}
where 
\begin{equation}
    \tilde\Delta= \Delta - 0.3422 \, V
\end{equation}
is an average, interface-configuration-independent energy gap, and the weak interface-configuration-dependent correction to the matrix element, accounted for by $\tilde\epsilon$, does not exceed a relative size of $0.2\%$.

For the blockade-violating process in Fig. 3b of the main text we can perform an analogous computation. We find
\begin{equation}
 \bigg(\frac{\Omega}{2}\bigg)^2
    \frac 1 {-\tilde\Delta^\prime + \tilde\epsilon^\prime}
\sum_j\big(\ket{\drmove}_{\; j}\bra{\urmove}+\ket{\urmove}_{\; j}\bra{\drmove}\big)
\end{equation}
where 
\begin{equation}
    \tilde\Delta^\prime=  2.2953 \, V - \Delta
\end{equation}
is an average, interface-configuration-independent energy gap, and the weak interface-configuration-dependent correction to the matrix element, accounted for by $\tilde\epsilon^\prime$, again does not exceed a relative size of $0.2\%$.

\bibliography{biblio}